# Experimental demonstration of the short bunch extraction by bunch rotation in a high-intensity rapid cycling proton synchrotron

Linhao Zhang[1,2,3,*] , Yukai Chen[1,3,*], Jingyu Tang[1,2,3,†], Xiao Li[1,3], Yang Liu[1,3], and Liangsheng Huang[1,3]

[1]*Institute of High Energy Physics, Chinese Academy of Sciences, Beijing, 100049, China*
[2]*University of Chinese Academy of Sciences, Beijing, 100049, China*
[3]*Spallation Neutron Source Science Center, Dongguan, 523803, China*

Short bunch proton beams are of great significance for the applications of white neutron beams and muon beams. The accelerator complex of China Spallation Neutron Source (CSNS) was designed to support the applications mainly based on neutron scattering techniques where the proton pulse length is not very sensitive. Some theoretical and experimental studies have been performed to see if one can extract a short-bunch proton beam by bunch rotation from the rapid cycling synchrotron (RCS) at CSNS. The experimental results at RCS have evidently displayed the bunch lengthening and rotation process, which demonstrates the effectiveness of this method even with a very short available time for the RF gymnastic processes and a high-intensity beam. With a beam power of 50 kW and normal longitudinal emittance at the injection, the proton beam with a bunch length of about 53% with respect to the one in the normal operation mode was obtained and transported to the spallation target. With a reduced longitudinal emittance at injection and the beam power of 30 kW, the shortest extraction bunch length obtained is about 26% of the one in the normal operation mode. Different machine settings have also been tested to show the impact of the desynchronization between the RF and magnetic fields, the influence of the non-adiabatic risetime and the adiabatic decay time of the RF voltage on the extraction bunch length. The experimental results are well consistent with the theoretical and simulated ones. It is interesting to observe that space charge has a beneficial effect on the bunch lengthening which will result in a shorter bunch at the extraction with the later bunch rotation. The controlled desynchronization method between the RF and magnetic fields in an RCS was also proven successful.

## I. INTRODUCTION

High power short pulsed proton beams play a mandatory role in producing secondary or tertiary beams such as neutron beams, gamma beams, muon beams, pion beams, neutrino beams, etc. For some research fields, experiments require pulsed beams with a shorter time duration, e.g. in a level of a few nanoseconds. However, most medium-energy proton synchrotrons adopt RF systems with relatively low frequency like a few MHz and deliver proton beams with a bunch length of tens of nanoseconds or even longer. Thus, the bunch compression method has been proposed to squeeze bunches before extraction in these accelerators. For example, at

* zhanglinhao@ihep.ac.cn, chenyk@ihep.ac.cn. These authors contributed equally to this work.
† Corresponding author. tangjy@ihep.ac.cn

China Spallation Neutron Source (CSNS) where the main applications are based on neutron scattering techniques, the rapid cycling synchrotron (RCS) delivers a proton beam of two bunches per pulse, and the full bunch length at the extraction is about 105 ns. However, the other experimental platforms, such as the Back-n white neutron source [1] and the EMuS muon source [2], demand shorter bunches. Therefore, a theoretical study to produce a proton beam with shorter bunches was carried out during the construction period of the CSNS facility [3].

Several methods can be employed to obtain shorter bunches in synchrotrons and storage rings. The most straightforward approach is to raise the RF voltage. However, the peak RF voltage from an RF system with a fixed number of cavities installed in the synchrotron is limited by the breakdown of the electric field, and it is not effective, as will be discussed in the next section. Another method is to add a higher harmonic RF together with the original one before the extraction to compose a dual-harmonic RF system [4], but the gain in bunch compression is still very limited, e.g. usually less than a compression factor of two with a new doubled frequency. The compression factor denotes the ratio of the original bunch length to the new bunch length. The method by applying the RF barrier technique is usually less efficient but probably good at dealing with space charge effects [5, 6]. The most effective method to attain short bunches that is widely accepted by the community is through the bunch rotation process [7-10]. This method first elongates the bunch by decreasing the RF voltage adiabatically or quickly shifting the RF phase to make the bunch situate at the unstable fixed point, and then rotates the bunch in the phase space either with a large RF voltage which rises very quickly in a more-or-less non-adiabatic way or by quickly shifting the RF phase back to the stable fixed point. The bunch rotation manipulation has been studied or practiced in slow-cycling synchrotrons or accumulator rings [10-12]. However, in a rapid cycling synchrotron, the beam acceleration that is dictated by a sinusoidal magnetic field is very fast. Thus, it is much more challenging to apply the method due to the very short available time for the RF gymnastics. Nevertheless, the early theoretical study demonstrates that a large bunch compression factor by the bunch rotation in an RCS is still possible with the help of the controlled desynchronization between the RF and magnetic fields [3].

This paper presents the first experimental results of bunch rotation in an RCS, which was conducted at CSNS/RCS. The controlled desynchronization method was also proven applicable. The structure of the paper is as follows: the introduction to the short bunch compression method and the experimental set-ups at CSNS/RCS in Sec. II, the experimental results of different scenarios and their comparison with the simulations in Sec. III, the experimental verification of the controlled desynchronization between the RF and magnetic fields in Sec. IV, and the conclusion in Sec. V.

## II. METHODS OF SHORT BUNCH EXTRACTION AND EXPERIMENTAL SET-UPS AT CSNS/RCS

### A. Methods of shortening the extraction bunch length

#### *1. Adiabatic bunch compression by raising RF voltage*

In small-amplitude approximation, the rms bunch length $\sigma_{\tau,i}$ in a matched RF bucket with voltage $V_0$ is [7]:

$$\sigma_{\tau,i} = \sqrt{\frac{A_{rms}}{\omega_0}\left(\frac{2|\eta|}{\pi heV_0\beta^2 E|\cos\phi_s|}\right)^{1/4}}, \tag{1}$$

where $A_{rms}$, $\omega_0$, $\eta$, $h$, $\phi_s$, $\beta$, and $E$ denote the longitudinal emittance in eV-s, angular revolution frequency of the reference particle, phase-slip factor, harmonic number, synchronous phase, Lorentz velocity factor, and total energy, respectively. Thus, the bunch length scales down with the fourth power of the RF voltage $V_0$ and can be shortened by raising $V_0$. However, this method is not very efficient with such low scaling power because the available RF voltage is limited by the breakdown of the electric field and the high cost of adding new RF cavities.

### *2. Non-adiabatic bunch rotation*

A more effective method called bunch rotation can be utilized by complex RF voltage manipulations to obtain short bunches. First, the RF voltage is decreased slowly or adiabatically from the initial value, $V_i$, to a very low one, $V_1$, during which the bunch gradually gets stretched following the change of RF bucket. Next, the RF voltage is quickly or non-adiabatically raised to a sufficiently high value, $V_2$, which is still below the maximum obtainable level by the RF system, then the mismatched bunch will perform rigid quadrupole mode oscillations and rotate in the newly established enlarged RF bucket. Finally, after 1/4 of the synchrotron oscillation period, the bunch becomes the shortest and is extracted immediately. The final shortest rms bunch length, $\sigma_{\tau,f}$, can be expressed by [3, 10]:

$$\sigma_{\tau,f} = \sqrt{\frac{A_{rms}}{\omega_0}\left(\frac{2|\eta|}{\pi h\beta^2 E|\cos\phi_s|}\right)^{1/4}}\frac{(eV_1)^{1/4}}{(eV_2)^{1/2}}. \tag{2}$$

Thus, the ultimate bunch compression ratio, as defined in Sec. I, can be expressed as:

$$r_c = \frac{\sigma_{\tau,i}}{\sigma_{\tau,f}} = \frac{V_2^{1/2}}{(V_0V_1)^{1/4}}. \tag{3}$$

Clearly, the bunch length now depends on the square root of the final RF voltage. Thus, the non-adiabatic bunch-rotation compression is much more efficient than the adiabatic bunch compression, but it is also more demanding for the transient response of the cavities and the low-level RF systems (LLRF).

However, the theoretical compression ratio by Eq. (3) is only achieved when the nonlinearity of the synchrotron frequency is small. With an elongated bunch covering a large fraction of the bucket, the bunch will get distorted during the rotation process because the outer region of the bunch suffers from a lower synchrotron frequency than the bunch core. This will happen when a very small lowest voltage $V_1$ is applied to stretch the bunch as possible. On the other hand, the minimum applicable $V_1$ is limited by the beam loading effect. A possibility to improve the bunch rotation is to linearize the RF voltage by adding a higher harmonic RF to the fundamental one, hence linearizing the synchrotron frequency of the particles [8,13]. In a rapid cycling synchrotron, the bunch compression will face a more severe nonlinearity problem because the bunch stretching and rotation processes are conducted with a non-zero RF phase. This will be solved by the controlled desynchronization method, as will be treated in detail in Sec. IV.

It is worth noting that the derivation of the theoretical compression ratio does not involve the beam intensity effects. The performance of the bunch rotation may be restricted by the longitudinal space charge effect as it counteracts the external RF focusing [14-16].

### *3. Reducing the initial longitudinal beam emittance*

As shown in Eq. (2), shorter bunches can be obtained by reducing the initial longitudinal beam emittance. In a high-intensity proton synchrotron with longitudinal painting, smaller longitudinal emittance can be achieved through special injection schemes: (a) Lower the RF voltage at the injection painting stage. However, the minimum voltage is restricted by beam loading and space charge effects which may cause beam loss. (b) Change the phase space painting in the longitudinal plane from off-momentum to on-momentum. (c) Increase the bunch chopping factor, defined as the ratio of the chopped-off duration to the chopper period, which is usually accompanied by the reduction of beam power without modifying the peak current in the linac injector. However, it is perfectly acceptable in our experimental study and for the future dedicated short-bunch extraction modes with very limited operation time at CSNS/RCS. When the longitudinal emittance is changed from the normal one, $A_{rms,i}$, to the one in short bunch extraction mode, $A_{rms,f}$, the corresponding bunch compression ratio is updated as:

$$R_c = \sqrt{\frac{A_{rms,i}}{A_{rms,f}}} \frac{V_2^{1/2}}{\left(V_0 V_1\right)^{1/4}} . \tag{4}$$

## B. Short bunch extraction in a rapid cycling synchrotron

The above bunch rotation method to obtain short extraction bunches has been successfully applied in compressor rings and slow cycling synchrotrons. However, the situation is quite different in a rapid cycling synchrotron, where one has to deal with the problems of beam acceleration during bunch compression and a very short available time slot for the RF gymnastics. Since the magnetic fields are driven by resonant power supplies and change in the form of a biased sinusoidal curve, the beam is still in a slow acceleration process before extraction. If the bunch compression process is conducted under this condition, the asymmetric RF bucket will distort the bunch and increase the final extraction bunch length. Therefore, the desynchronization between the RF and the magnetic fields was proposed to create a flattop phase before extraction at an RCS [3], which assists bunch compression to be performed in a stationary and symmetrical RF bucket. Moreover, the extraction moment can be delayed slightly to provide as much time as possible for executing the complicated RF gymnastics. The closed-orbit distortion due to the desynchronization is inevitable because of the momentum difference from the reference one. However, the distortion can be mitigated by keeping the desynchronization moment and extraction moment approximately symmetric (before and after) with respect to the moment of the peak magnetic field. Since the two moments are close to the peak magnetic field, the closed-orbit distortion is relatively small and considered acceptable.

## C. Beam properties and experimental settings at CSNS/RCS

The CSNS accelerator complex comprises an 80-MeV H- linac and an RCS, both working at a repetition rate of 25 Hz. The RCS is a four-fold symmetric ring with a circumference of 227.92 m, which accelerates the proton beam with two accumulated bunches to 1.6 GeV from 80 MeV. Then the beam is extracted and transferred to the spallation target to produce neutrons. After the beam commissioning and gradual power ramp-up in about two years, the RCS reached the target beam power of 100 kW in February 2020. The longitudinal dynamics, including the synchronicity between the ramping magnetic field and the RF system and the mitigation of the space charge effects, was carefully treated in the commissioning to minimize beam losses [17]. The main parameters of CSNS are shown in Table I.

TABLE I. Main Parameters of the CSNS accelerator complex at Phase I.

| | |
|---|---|
| RCS extraction beam power (kW) | 100 |
| RCS injection/extraction energy (GeV) | 0.08/1.6 |
| RCS circumference (m) | 227.92 |
| RCS harmonic number | 2 |
| RCS transition gamma | 4.896 |
| RCS injection/extraction RF frequency (MHz) | 1.0241/2.444 |
| Pulse repetition rate (Hz) | 25 |
| Protons per pulse ( $10^{13}$) | 1.56 |
| Linac macro pulse peak current (mA) | 15 |
| Linac rf duty factor (%) | 1.5 |
| Linac beam momentum spread (rms) | <±0.1% |

In the normal operation mode of the CSNS/RCS with the beam power of 100 kW, the macro pulse length of 415 μs in the linac with a chopping factor of 41 % is injected into the RCS. As the chopping frequency is exactly equal to the injection RF frequency of 1.0241 MHz, two equidistant bunches are formed in the RCS. The injection energy is about 80.20 MeV, which is a little deviated from the one corresponding to the injection RF frequency to perform the so-called off-momentum painting in the longitudinal phase space as one of the measures to mitigate the space charge effects. The injection RF voltage is about 36 kV, which offers a large enough RF bucket to accommodate the injection beam with longitudinal painting. The injection process starts at -0.15 ms, referring to zero for the moment of the minimum magnetic field, and the total injection turn is about 210. After the injection and RF capture processes, the two bunches are accelerated from 80 MeV to 1.6 GeV during 20 ms to follow the sinusoidal magnetic field. Finally, a group of eight extraction kickers are triggered to extract the two bunches at 19.80 ms. The corresponding RF voltage is 60 kV at the extraction. The RF voltage and frequency patterns used during the acceleration cycle are depicted in Fig. 1.

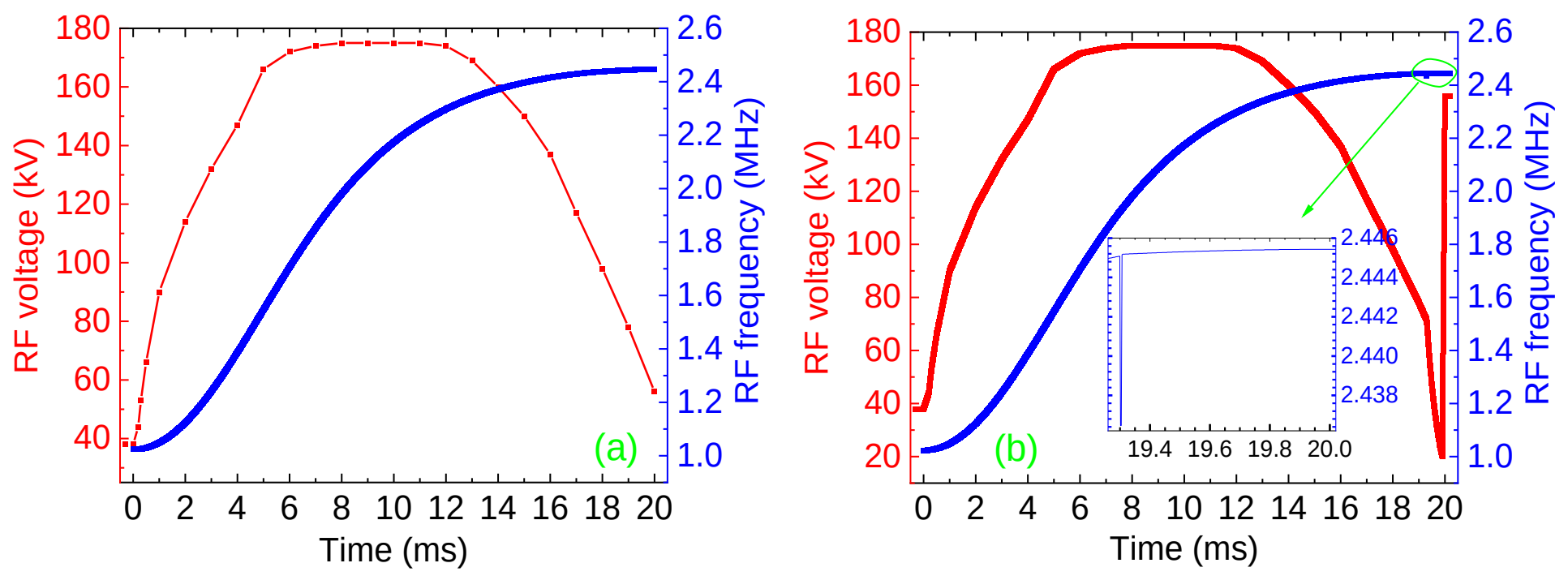

FIG. 1. RF voltage and frequency patterns: (a) the normal operation mode; (b) the short bunch extraction mode. The RF voltage and frequency patterns before extraction are modified to accomplish the bunch compression and the desynchronization between the RF and magnetic fields (see Sec. IV).

The beam properties, like bunch profile and bunch phase, can be measured by various beam instruments. The wall current monitor (WCM) with a bandwidth of 200 MHz in the RCS is employed to measure the turn-by-turn bunch profile within 25 ms, which covers the duration from beam injection to extraction. Besides, the bunch length for the extracted beam can be acquired from the current transformer (CT02) in the RTBT (Ring to Target Beam Transport) beamline. The bunch center phase can be obtained from the fast current transformer (FCT) in the RCS. The complex digital low-level RF system controls the RF voltage and frequency according to the programmed input values, as shown by Fig. 1. There are 32 beam position monitors (BPM) spreading throughout the RCS to measure the beam orbit information, including the closed-orbit errors that are important for monitoring the synchronization among the beam momentum, the RF frequency and the magnetic fields during acceleration.

In the machine studies for short bunch extraction, the CSNS accelerator complex was operated in the repetition frequency of 1 Hz unless otherwise stated, for the safety reason to operate such a high beam power accelerator. The two bunches were extracted in a single turn to the beam dump (R-dump) located at the RTBT. Different beam powers can be achieved by modifying the chopping factor and the peak current through the linac front-end. As an example shown in Fig. 1(b), the adopted RF voltage before extraction is slowly lowered from 71.4 kV at 19.3 ms to the minimum of 20 kV at 19.9 ms, then quickly raised to 156 kV with the non-adiabatic risetime of 100 μs and then keeps unchanged until the extraction moment at 20.2 ms. The RF frequency was also modified to implement the controlled desynchronization between the RF and magnetic fields, which is explained in detail in Sec. IV.B.

## III. EXPERIMENTAL RESULTS AND COMPARISON WITH THE SIMULATIONS

### A. Analysis of beam monitor signals

In the daily operation and machine studies of CSNS/RCS, the bunch length during the whole acceleration cycle can be measured turn-by-turn by the WCM, as shown in Fig. 2(a). The larger beam signal amplitude indicates the shorter bunch. The bunch signal at the second half of the acceleration cycle fits well with a Gaussian distribution. After the Gaussian fitting of the bunch signal from 15 ms to 19.8 ms or the extraction moment, the corresponding rms bunch length can be obtained, see Fig. 2(b). One can clearly see the bunch length oscillation, i.e., the quadrupole oscillation, which arises from the mismatch between the bunch shape and the RF bucket distortion caused by space charge. The rms extraction bunch length is about 26.5 ns in this normal operation mode.

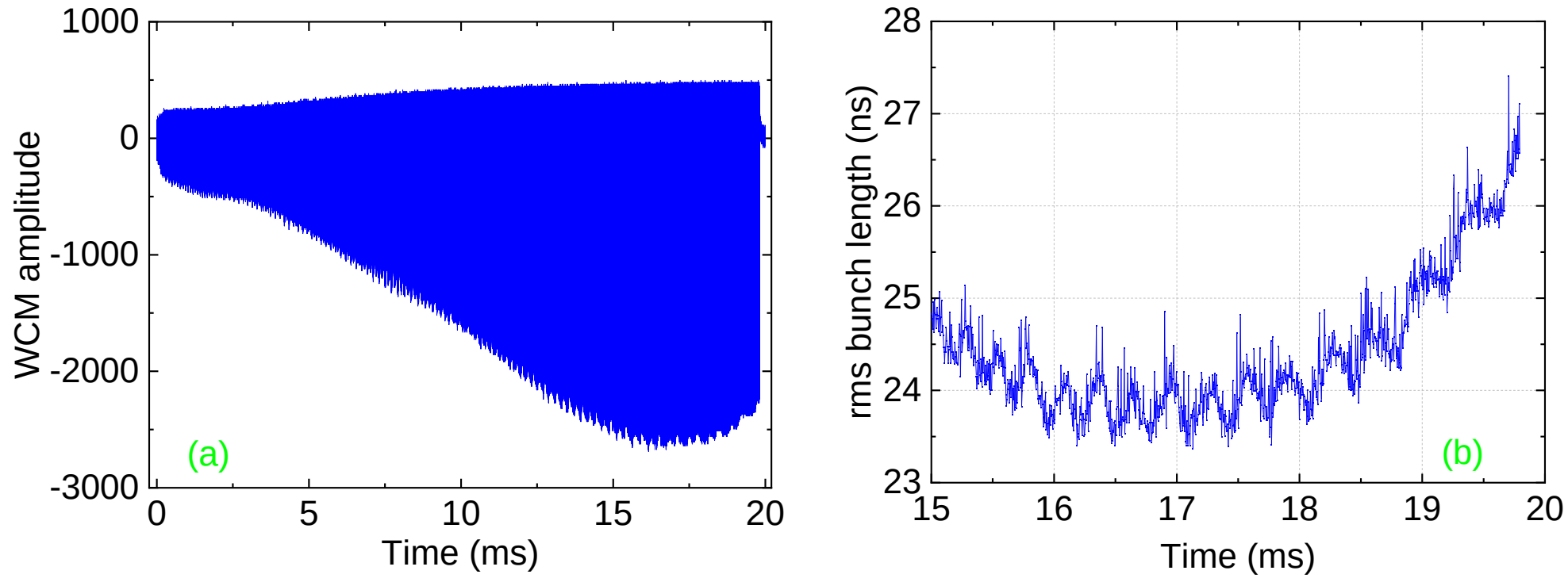

FIG. 2. (a) The beam signal during the whole acceleration cycle measured by the WCM. (b) The rms bunch length from 15 ms to 19.8 ms or the extraction moment with the corresponding bunch signal fit by a Gaussian function.

Figure 3(a) presents the charge distribution of the two bunches in the RCS ring from 15 ms to 19.8 ms, which is analyzed from the original WCM signal. One can observe the evident oscillation of the bunch center, i.e., the dipole oscillation, whereas the quadrupole oscillation is too small to be visible. Figure 3(b) further shows the dipole oscillation of the phase of the bunch center that is measured by the FCT. Compared with Fig. 2(b), it is evident that the oscillation period of the bunch length measured by the WCM is twice that of the bunch phase measured by the FCT. This is easy to understand because the quadrupole oscillation is a modulation of bunch length at twice the synchrotron frequency, while the dipole oscillation is at the synchrotron frequency. Thus, the longitudinal motion in the CSNS/RCS demonstrates the combination of dipole and quadrupole oscillations.

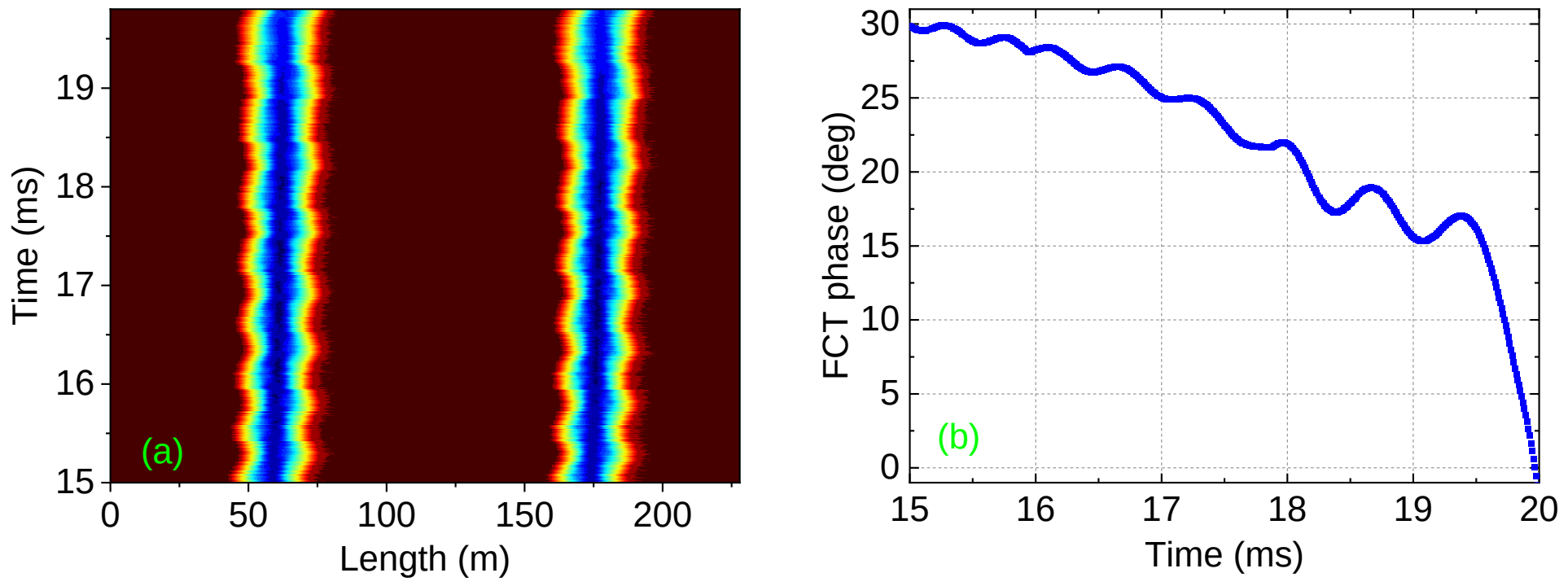

FIG. 3. (a) Charge distribution of the two bunches in the CSNS/RCS measured by the WCM from 15 ms to 19.8 ms. (b) Oscillation of the bunch center measured by the FCT in the same period.

Figure 4 depicts the comparison of the rms bunch length evolution from 10 ms to the extraction among the WCM measurement, the theoretical calculation, and the simulation using the ORBIT code [18] in the normal operation mode. One can see that they agree very well. However, it is worth pointing out that the ORBIT code has been revised to keep the longitudinal emittance in eV-s invariant during acceleration when applied to an RCS [19]. Besides, in order to make the ORBIT simulation results consistent with the experimental and theoretical ones, the RF voltage in the simulation was 1.1 times as the designed ones, the chopping factor in the linac was changed from 41% to 33.3%, and the injection kinetic energy was modified from the original 80.2 MeV to 80.1 MeV. These modifications are considered acceptable since they are within the parameter fluctuation error ranges in the practical accelerator operation.

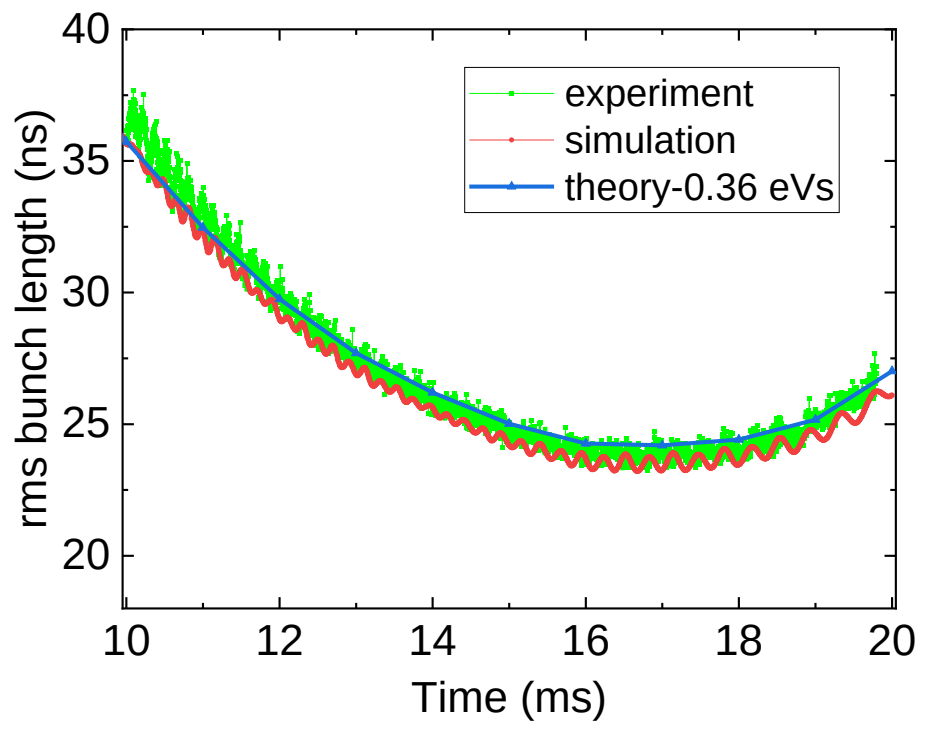

FIG. 4. Evolution of the rms bunch length from 10 ms to the extraction from the measurement (green), the theoretical calculation (blue), and the ORBIT simulation (red). The theoretical bunch length is calculated based on Eq.(1) with the rms bunch area of 0.36 eVs.

### B. Experimental demonstration of the bunch-rotation compression

After the machine studies have shown the bunch-rotation process and its effectiveness with different set-ups, the so-called target-shooting experiment was conducted to demonstrate further its application in a practical operation mode, in which CSNS worked with the repetition rate of 25 Hz and a beam power of 50 kW and the extracted beam was transported to the spallation target. This target-shooting experiment lasted around one hour very reliably, and the lower beam power than the usual 100 kW was to avoid frequent triggers for the machine protection system due to beam losses in higher beam power that will be solved in the future. This beam power was achieved without changing the chopping factor but with reducing the linac peak current by adjusting the front-end, which means the longitudinal emittance in the RCS did not change. The RF voltage and frequency follow the patterns shown in Fig. 1(b) to accomplish the bunch compression and the controlled desynchronization between the RF and magnetic fields, respectively. The beam extraction moment is shifted from 19.80 ms in the normal operation mode to 20.08 ms, at which moment the bunch is shortest.

The extraction kickers and the correctors in the RTBT needed to be only slightly adjusted from the 100-kW normal operation mode, which indicates that the closed orbit deviation caused by the desynchronization is very small. Additionally, the beam loss pattern in the ring was almost the same as the usual one, as shown in Fig. 5, which further illustrates that this operation mode can be applied safely.

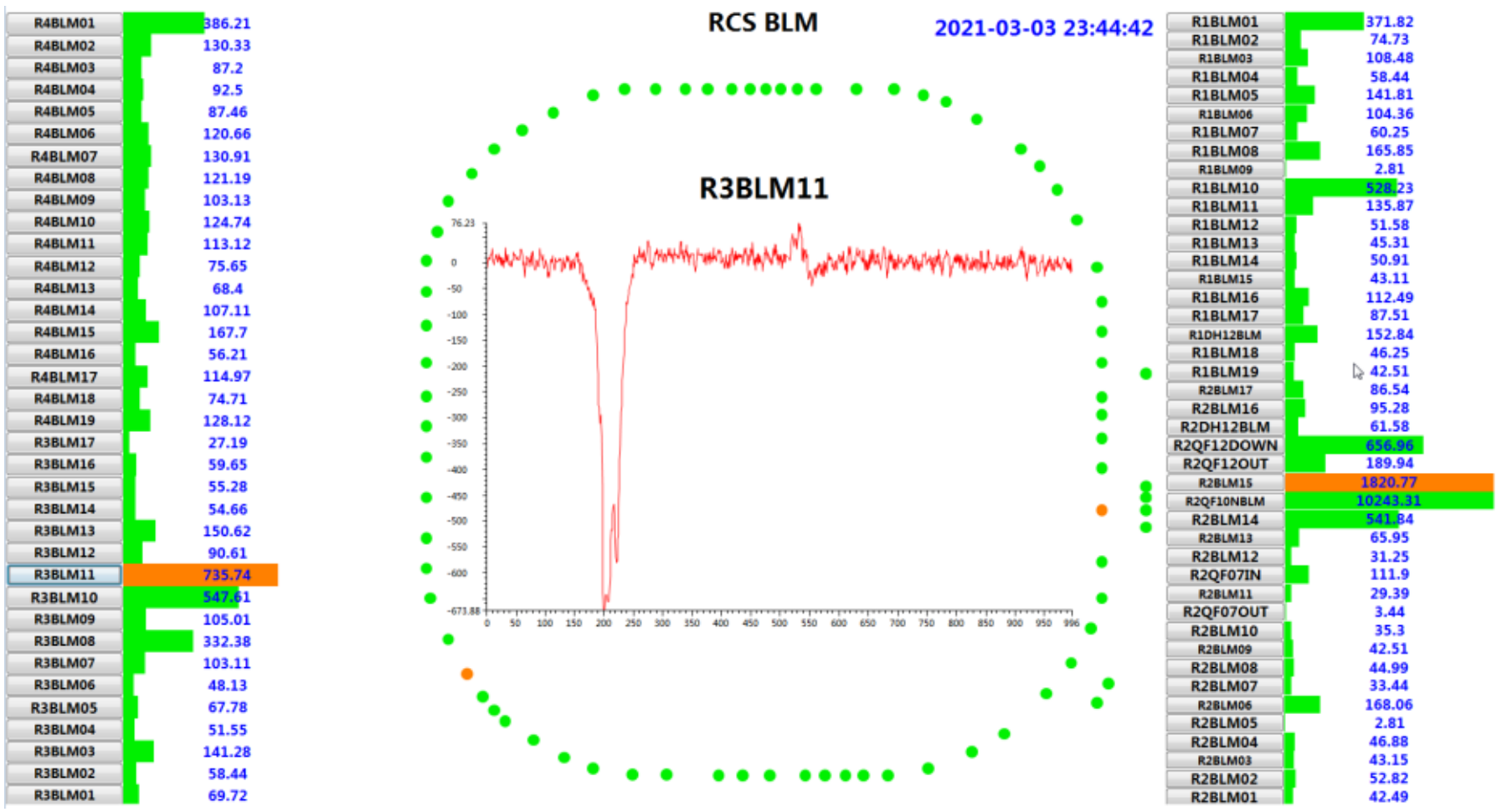

FIG. 5. Beam loss distribution across the whole RCS measured by beam loss monitors (BLM) in the target-shooting mode. The green color indicates that the beam loss is totally negligible, and the orange color gives a warning of beam loss level but still below the threshold to trigger the machine protection system.

Figure 6(a) presents the charge density distribution of the two bunches from 19.3 ms to the extraction. One can clearly see the whole bunch compression process that includes the adiabatic bunch lengthening and the fast bunch rotation. Figure 6(b) further gives the evolution of the rms bunch length in the period. The rms bunch length at the extraction is 14.8 ns, which is about 53.1% of the one in the normal operation mode. The ratio is slightly larger than the theoretical one, i.e., 47.1%, calculated by Eq. (3). This is mainly due that the actual RF voltage did not reach the designed lowest value of 20 kV during the voltage manipulation, which is caused by the lagging effect with a rapid change of the voltage. Besides, the very short duration of the lowest voltage causes insufficient adiabatic bunch stretching due to the slow synchronous oscillation and thus leading to a slightly larger bunch length after rotation.

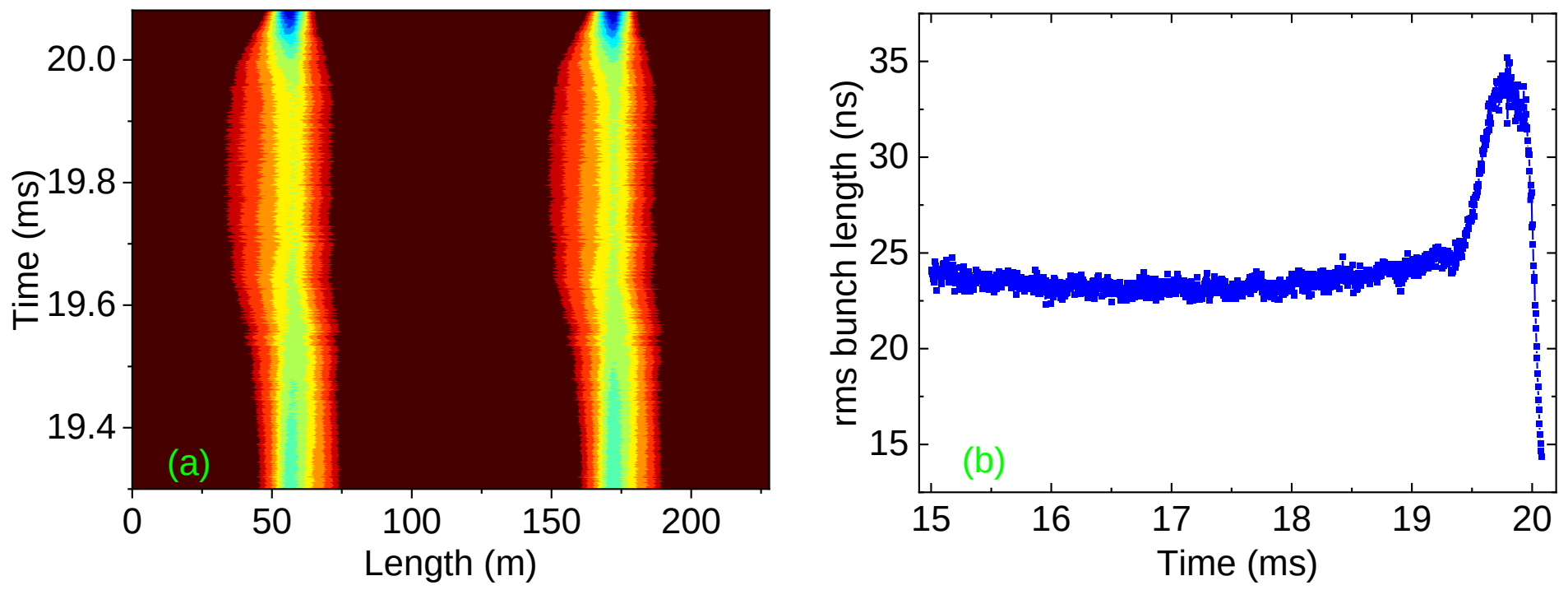

FIG. 6. Evolution of the charge distribution of the two bunches in the RCS from 19.3 ms to the extraction (a) and the evolution of the rms bunch length from 15 ms to the extraction (b) in the target-shooting mode.

The short-bunch proton beam extracted from the RCS was also measured by CT02 in the RTBT, which gives the consistent rms bunch length from the WCM measurement. Additionally, the pulse width of the proton beam was checked indirectly by measuring the γ-flash from the spallation target that has a slightly widened bunch length as compared to the primary proton beam due to the target thickness, see Fig. 7. This was carried out by using a fast-response plastic scintillation detector placed in the back-streaming neutron beamline (the Back-n WNS) [1]. The Gaussian fit to the first peak in Fig. 7 gives 28.4 ns (rms) for the normal operation mode and 15.7 ns (rms) for this target-shooting mode. The baselines at the second peaks are levelled up by the high-energy neutrons from the first proton bunch on target.

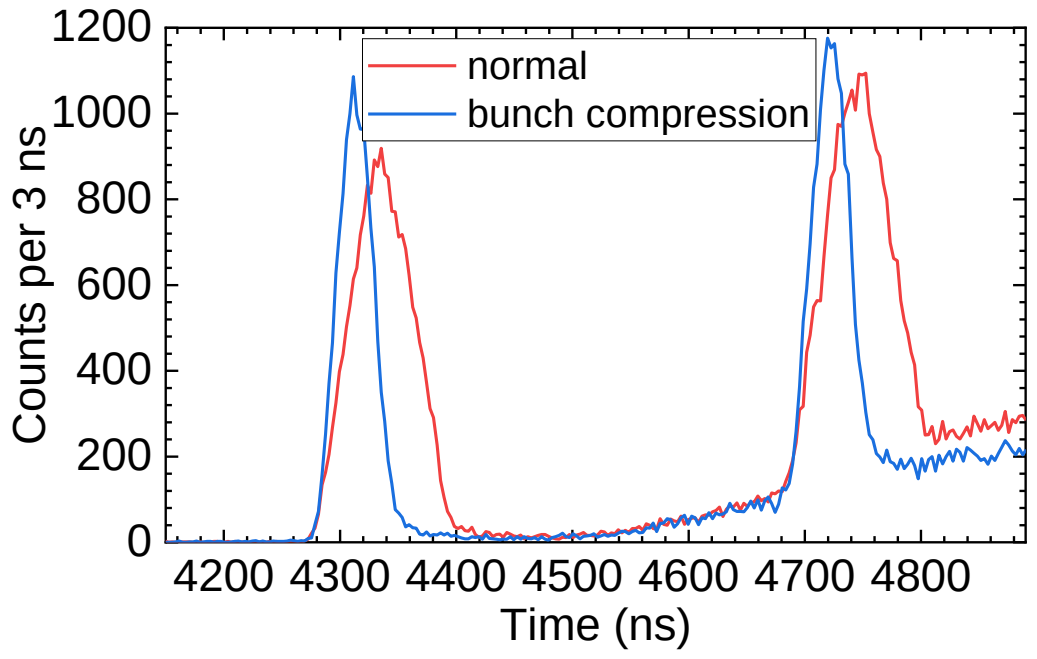

FIG. 7. Proton bunch length by measuring the γ-flash response waveform in the Back-n white neutron beamline: the normal operation mode (red); the target-shooting mode with the bunch compression (blue). It shows that the proton bunch length in this target-shooting mode is approximately half of that in the normal mode.

### C. Bunch compression with different longitudinal emittance at injection

As a smaller longitudinal emittance is helpful in obtaining shorter bunches at extraction no matter if the bunch compression method is applied, different longitudinal emittances at injection have been tested to verify this. The machine studies were conducted at the equivalent beam power of 30 kW at 25 Hz and with different chopping factors in the linac that correspond to different longitudinal emittances, as explained in Sec. II.A.3. The peak current in the linac was adjusted to follow the change in the chopping factor to maintain the same beam power. The RF voltage and frequency patterns are almost the same as the setting of Fig. 1(b), except that the minimum voltage is set as 15 kV. Table II shows the comparison among the extraction bunch lengths with different chopping factors. The shortest extraction bunch length by experiment that corresponds to the chopping factor of 75% is about 26% of the one in the normal operation mode.

TABLE II. Comparison on the extraction bunch length among the different chopping factors.

| Chopping factor | RMS bunch length (ns) | | | Bunch length with respect to the normal mode | | |
|---|---|---|---|---|---|---|
| | Experiment | Simulation | Theory | Experiment | Simulation | Theory |
| 55% | 10.67 | 9.55 | 9.43 | 40.3% | 36.2% | 35.4% |
| 60% | 9.94 | 8.88 | 8.7 | 37.5% | 33.6% | 32.7% |
| 65% | 9.26 | 8.63 | 8.14 | 35.0% | 32.7% | 30.6% |
| 70% | 8.69 | 8.07 | 7.78 | 32.8% | 30.6% | 29.2% |
| 75% | 6.85 | 7.75 | 7.41 | 25.6% | 29.4% | 27.8% |

The changing trend of the extraction bunch length obtained by experiment, simulation, and theory under different chopping factors is depicted in Fig. 8(a). Evidently, the greater chopping factor will lead to the shorter final extraction bunch. Typically, the extraction bunch obtained by experiment is longer than that obtained by theory and simulation, since the RF voltage in the experiment cannot fully reach the design value and there are various errors and beam intensity effects in practice. However, the extraction bunch under the chopping factor of 75% is even shorter than the theoretical one. It can be explained by that with the same beam power a larger chopping factor will lead to the higher peak current in the RCS due to the smaller longitudinal emittance. Thus, the space charge effects become stronger in the acceleration period prior to the RF gymnastics process and render the dramatic mismatch between the bunch shape and the RF bucket, which causes the oscillation of bunch length, see Fig. 8(b). Fortunately, this quadrupole oscillation can be profited to enhance the bunch lengthening process if the non-adiabatic rising moment starts from when the bunch length oscillation reaches the maximum, which was the case of 75% chopping factor by coincidence. This issue will be further studied in Sec. III.F.

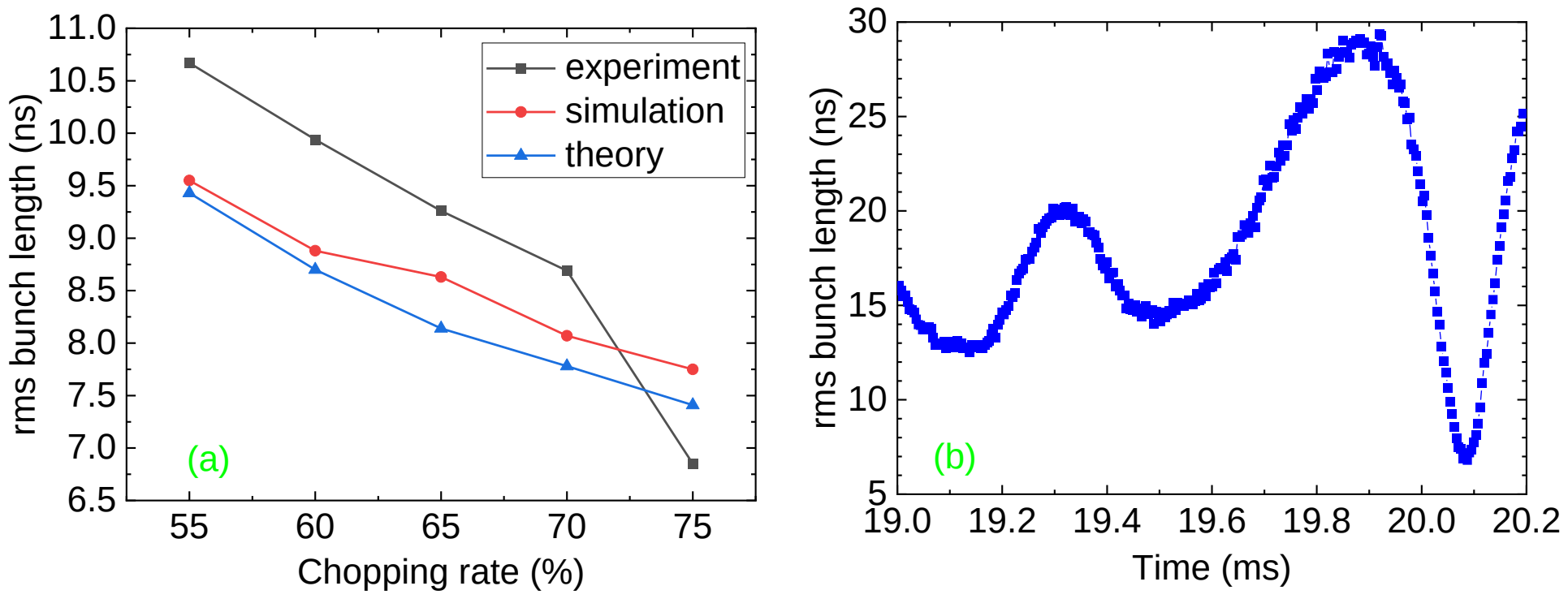

FIG. 8. (a) The changing trend of the extraction bunch length obtained by experiment, simulation, and theory under different chopping factors. (b) The effect of the dramatic quadrupole oscillation caused by the strong space charge under the chopping factor of 75% on the process of the whole bunch compression.

### D. Influence of the non-adiabatic risetime of the RF voltage

Ideally, the non-adiabatic risetime of the RF voltage for the bunch rotation should be as short as possible, at least within 100 μs or about one-fifth of the synchrotron period at the top voltage. However, the RF systems, including the ferrite-loaded RF cavities, the power sources, and the LLRF, are not so easy to ramp up the voltage very quickly. After many efforts to modify the LLRF and the power sources, the total RF voltage with all the eight cavities can be raised from 20 kV to the maximum 160 kV in less than 100 μs, with the smallest 50 μs. Some tests with different risetimes were performed to see the effect on the bunch compression. The chopping factor of 60% and the equivalent beam power of 30 kW at 25 Hz, and the RF voltage from 15 kV at 19.9 ms to 156 kV with 50 μs, 75 μs, and 100 μs, were applied. The results are compared in Fig. 9, which also includes the simulation results. One can see that the simulation results agree well with the experimental ones, both of which indicate that a shorter risetime will lead to a shorter bunch length but the influence is almost insignificant from 50 μs to 100 μs.

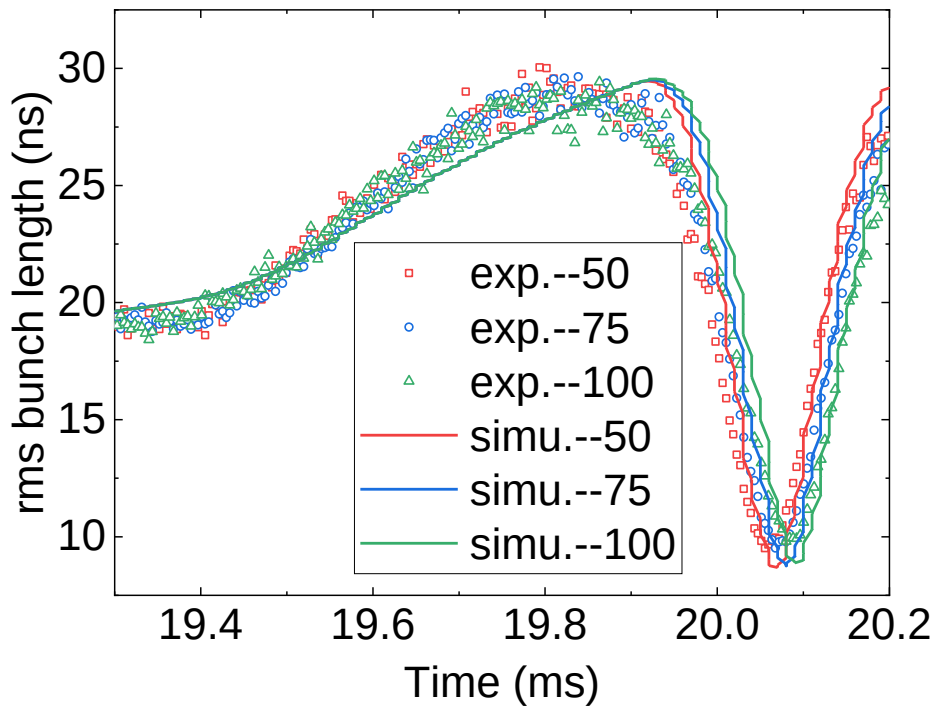

FIG. 9. Measured and simulated evolution of bunch length during the bunch stretching and bunch rotation for three RF voltage risetimes of 50 μs, 75 μs and 100 μs.

### E. Influence of the adiabatic decay time of the RF voltage

In accumulator rings or slow-cycling synchrotrons, usually, the time to decrease the RF voltage slowly to stretch the bunch adiabatically can be satisfied easily. However, with a rapid cycling synchrotron, it does pose problems. Some machine studies were performed to investigate the effect of the adiabatic decay time of the voltage on the bunch compression. There the RF voltage pattern was designed to study the bunch lengthening process only without considering the non-adiabatic bunch rotation process, but the controlled desynchronization was still applied. As shown in Fig. 10 (a), the RF voltage was decreased exponentially from 71.4 kV at 19.3 ms to 15 kV with a decay time of 0.5 ms, 0.6 ms, and 0.7 ms, respectively, and then remained unchanged until extraction at 20.20 ms. The same injection condition as in Sec. III.D was adopted with the chopping factor of 60% and equivalent beam power of 30 kW at 25 Hz.

Figure 10(b) shows the evolutions of bunch length for the different decay times by experiments and simulations. One can see that the trends of the simulation results and the experimental ones are in good agreement. Both of them indicate that the combination of a shorter decay time, e.g. 0.5 ms, and a flat-bottom time of 0.2 ms will be more conducive to obtaining longer bunches during the bunch stretching process. From the simulations, one can find that the bunch is slightly tilted at the lowest RF voltage due to extremely slow synchrotron motion, the bottom time helps the bunch to stretch further by rotation.

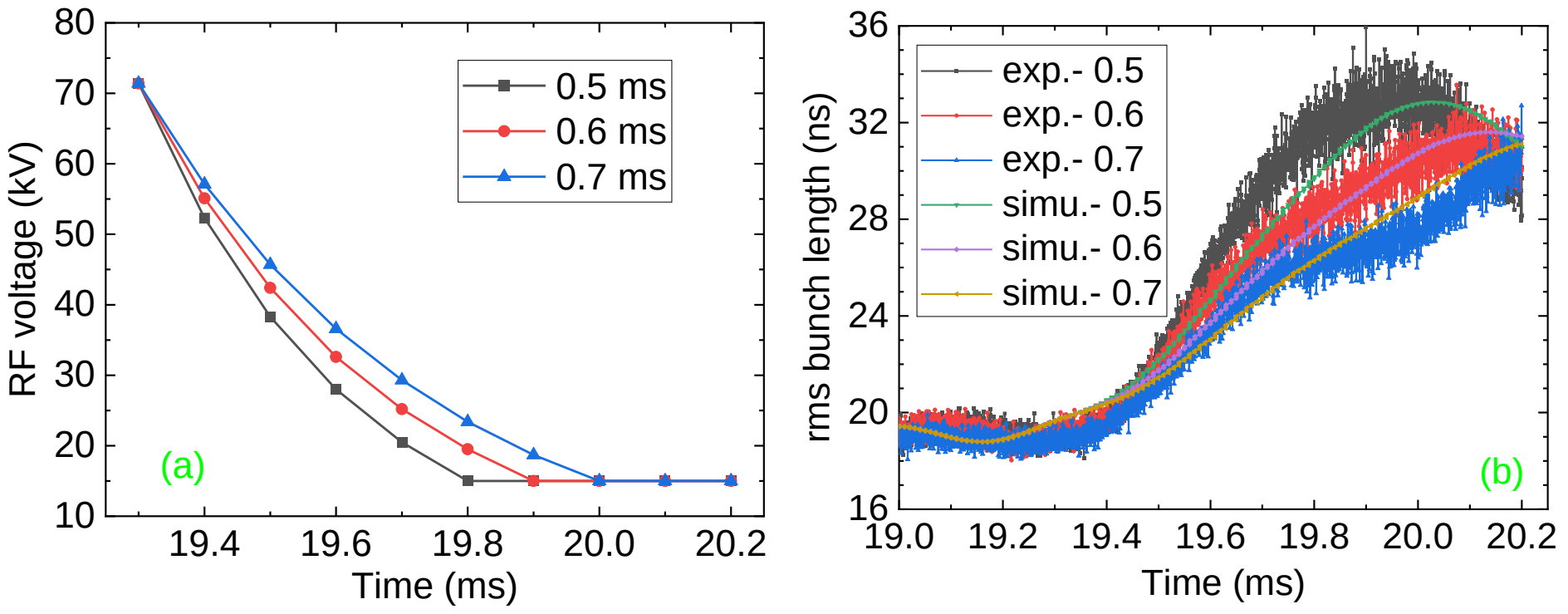

FIG. 10. (a) The RF voltage curves for different decay times: 0.5 ms, 0.6 ms and 0.7 ms. (b) The evolutions of the bunch length during the bunch stretching process from both the experiments and simulations.

### F. Influence of space charge on the bunch lengthening process

The influence of the quadrupole oscillation caused by space charge on the extracted bunch length was mentioned in Sec. III.C, where the oscillation feature of bunch length was profited to extract shorter bunches. Here, more detailed study is given, including both the adiabatic bunch lengthening and non-adiabatic bunch rotation processes. Figure 11 shows the simulation results about the space charge effect on the bunch stretching and bunch rotation. One can clearly see that the extraction bunch is shorter with space charge, which indicates that the beneficial effect of space charge on the adiabatic bunch stretching process overtakes its detrimental impact on the bunch rotation process. As a comparison, previous studies in other machines tend to focus on the bunch rotation stage [14-16] because the bunch becomes shorter and the bunch density becomes higher in this stage, which will lead to the stronger space charge effects and the possible bunch distortion or a longer bunch length after a 90° bunch rotation.

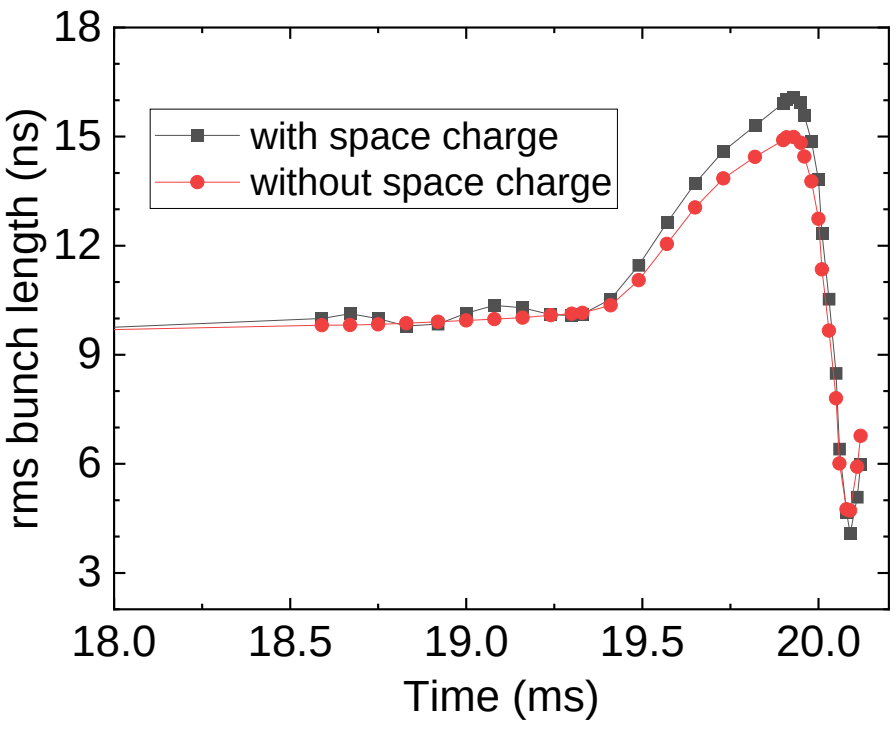

FIG. 11. Simulation results about the effect of space charge on the bunch length during the bunch stretching and bunch rotation processes. Simulation conditions: beam power of 30 kW and chopping factor of 70%.

# IV. IMPLEMENTATION OF THE CONTROLLED DESYNCHRONIZATION METHOD

## A. Principle of the controlled desynchronization between the RF and magnetic fields

In an ordinary synchrotron, the magnetic field changes with the beam energy. Besides, the RF frequency is an integer multiple (usually called harmonic number) of beam revolution frequency, which means that the RF frequency is also linked to the beam energy. Therefore, the RF frequency, $f_{rf}$, is synchronized with the magnetic field, $B$, as expressed below:

$$f_{rf}(t)=\frac{hc}{C}\left[\frac{B^2(t)}{B^2(t)+(m_0c^2/qc\rho)^2}\right]^{1/2}. \qquad (5)$$

Here, $C$ is the circumference of the synchrotron, $c$ is the speed of light, $m_0$ and $q$ are the rest mass and charge of the particle, respectively, $\rho$ is the curvature radius of the bending magnet.

However, in a rapid cycling synchrotron, the magnetic field is driven by a resonant power supply and changes in the form of a sinusoidal curve with a bias field, so the beam is still in a slow acceleration just before extraction. If the bunch compression by a complex RF gymnastics is conducted under this condition, the asymmetric RF bucket will distort the bunch and further increase the final bunch length. In addition, the bucket size is smaller with non-zero RF phase for the same RF voltage, which is not favored for both the bunch stretching process and the bunch rotation of a long bunch. Therefore, the method of controlled desynchronization between RF and magnetic fields was proposed to overcome these problems [3]. The main measure is to quickly move the RF phase to the bunch center before implementing the bunch compression process, and then the beam will become stationary as in an accumulator ring while the magnetic field is still rising slowly. If the desynchronization starts at a time not far from the extraction where the magnetic field reaches the maximum or 90° in the sinusoidal curve, e.g. 0.5 ms for an RF period of 40 ms or 4.5° in the RF phase, the error in the closed orbit caused by the desynchronization due to the momentum deviation from the reference will be acceptable.

The operation of rapidly shifting the RF phase is practiced in many accelerators, particularly for the RF manipulation to cross the transition energy [20, 21]. The time required for the phase jump of less 10° in the CSNS/RCS is estimated to be a few microseconds.

### B. Simulations of the controlled desynchronization

The input RF frequency curve of the CSNS/RCS for the normal operation mode, as shown in Fig. 1, includes 8192 frequency points evenly distributed from 0 to 20 ms in an LLRF file. The frequencies before 0 ms and after 20 ms are set as the frequencies at 0 ms and 20 ms, respectively.

To provide simulation guidance for the desynchronization between RF and magnetic fields, a one-dimensional longitudinal tracking code was developed according to the input parameters of CSNS/RCS [19]. Additionally, a two-step method of desynchronization is proposed to be implemented in the machine. Taking the desynchronization moment at 19.3 ms as an example, the first step is to shift the RF phase to zero by adding a sudden change to the two frequency points near 19.3 ms, which can be calculated by dividing the RF phase to shift by the duration between the two frequency points, i.e., 4.88 μs. The second step is to match the frequency following 19.3 ms to maintain the bunch phase to zero degree or oscillate slightly around zero degree. Because the magnetic field is still ramping up at this moment, the path of the particles traveling the RCS will decrease due to no-zero momentum compaction. To maintain the bunch phase at zero, the subsequent RF frequency will be slightly increased, as shown in Fig. 12(a). Figure 12(b) presents the simulated bunch phase after adopting the desynchronization frequency curve. One can see that the bunch phase after 19.3 ms is zero. It is worth mentioning that the well-matched desynchronization frequency curve is weakly linked to the voltage, and a small phase oscillation around zero caused by different voltage patterns is entirely acceptable.

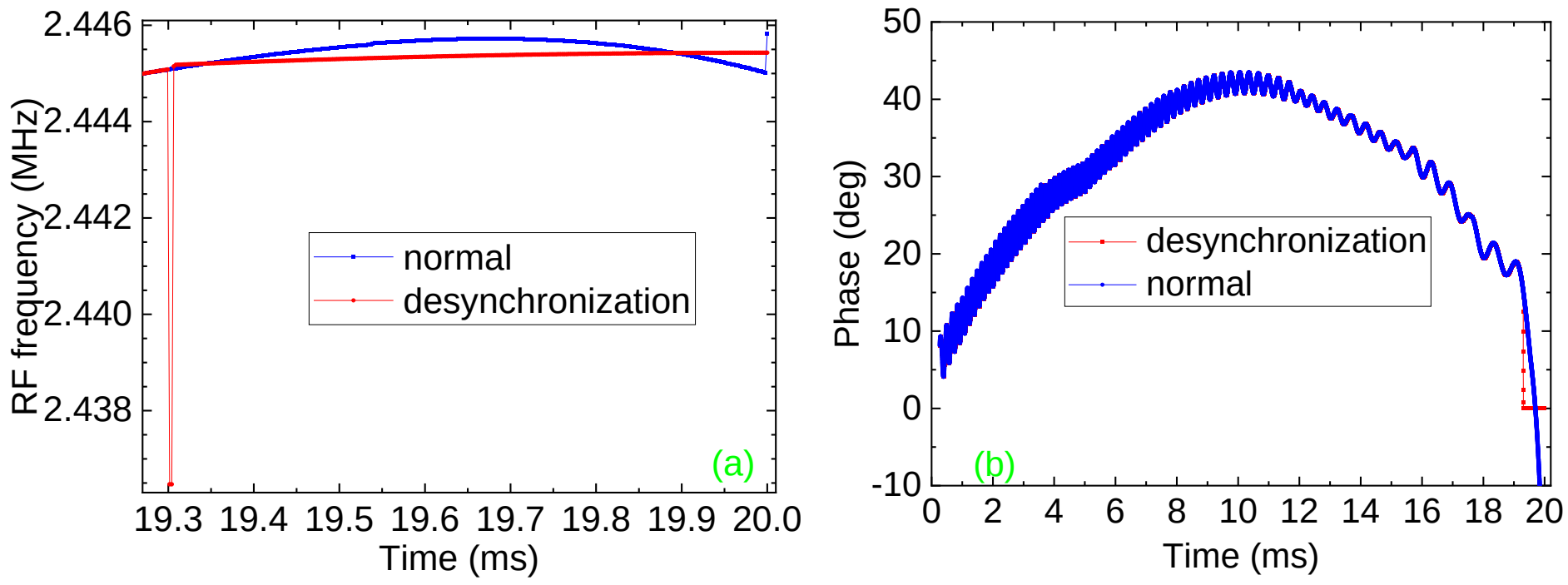

FIG. 12. (a) The RF frequency pattern for the normal mode (blue) and the desynchronization mode obtained from the simulation (red). (b) The simulated bunch phase during the CSNS/RCS acceleration cycle for the normal mode (blue) and the desynchronization mode (red).

### C. Experimental verification of the controlled desynchronization

We experimentally verified the controlled desynchronization process between the RF and magnetic fields by analyzing the evolutions of the amplitude of the WCM signal, the FCT phase, and the BPM beam centroids, as shown in Fig. 13. From the perspective of the WCM signal, see Fig. 13(a), the bunch signal is basically symmetric with the desynchronization before and after the maximum amplitude where the shortest bunch reaches after a 90° non-adiabatic rotation, which indicates that the bunch is symmetric and rotates around the bucket center in the bunch rotation process. As a comparison, without executing the desynchronization, the bunch is asymmetric due to the non-symmetric bucket with a non-zero phase when the bunch rotation starts, as shown by the WCM signal. Figure 13(b) represents the bunch phase oscillation with the desynchronization from the FCT phase. After the desynchronization moment at 19.3 ms, the phase oscillation is within ±3°, while a much larger phase off-centering is observed for the case without applying the desynchronization. The BPM beam centroid is an averaged value over three BPMs symmetrically located in three arc regions with a large dispersion, which corresponds to the momentum deviation from the reference one defined by the magnetic field. Figure 13(c) shows a clearly increased closed-orbit deviation after the moment of applying desynchronization as compared with the usual one, but it is well under control as expected.

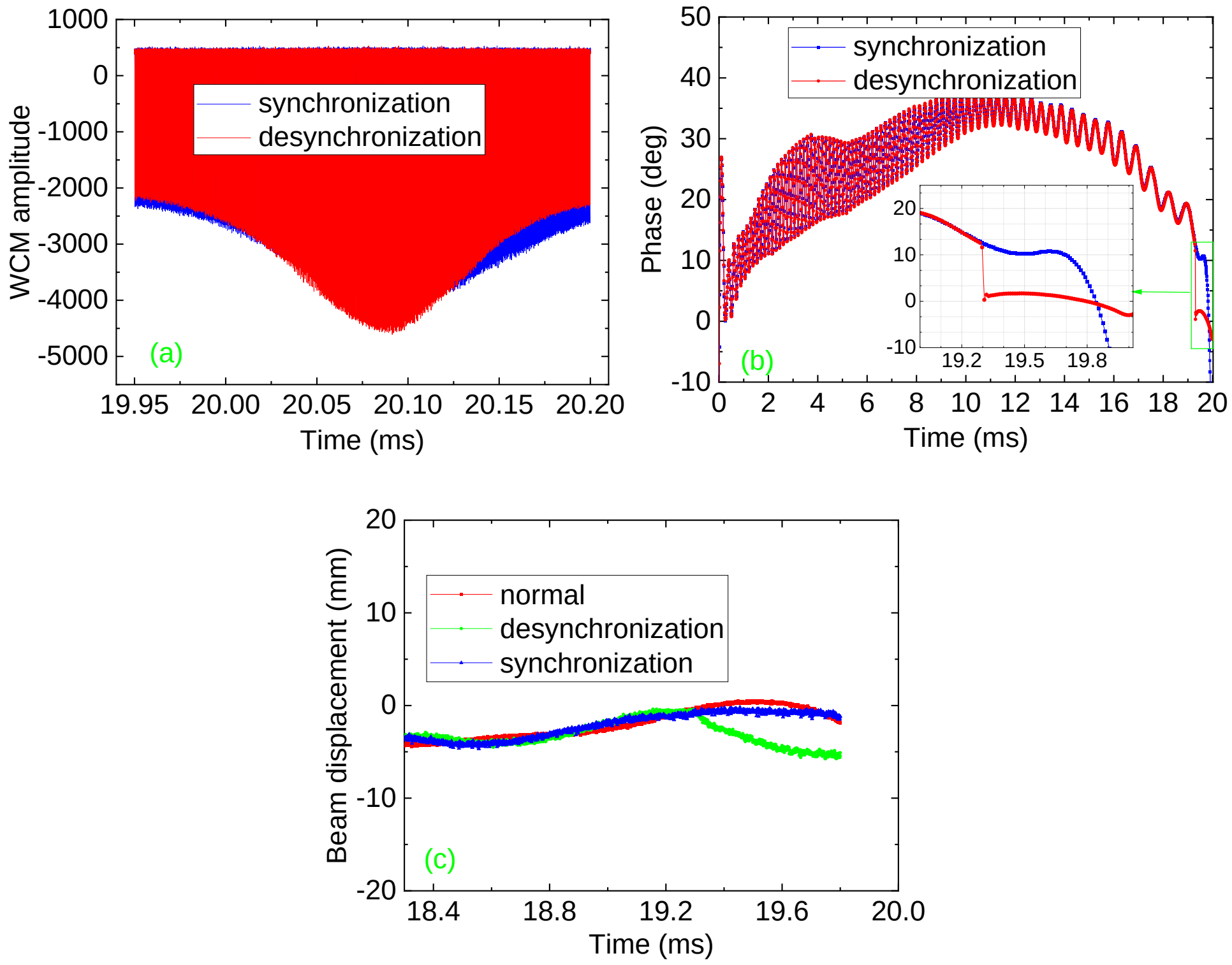

FIG. 13. Experimental verification of the controlled desynchronization from: (a) the analysis of the amplitude of the WCM signal; (b) the FCT phase; (c) the BPM beam centroid. "synchronization" denotes that only the RF voltage pattern is applied to extract short bunches, without modification to the RF frequency and phase. Experimental conditions: beam power of 30 kW and chopping factor of 60% for (a) and (b); beam power of 50 kW and chopping factor of 41% for (c).

### D. Controlled desynchronization in different machine settings

The controlled desynchronization method was applied for all the above machine studies, and its effect on attaining short extraction bunches in different cases was studied systematically, and compared with the "synchronization". Three types of beam power of 30 kW, 50 kW and 100 kW were chosen, with the similar injection conditions mentioned before. The chopping factor for the 30-kW case was 60% to have a smaller longitudinal emittance, whereas the original 41% was applied for the cases of 50 kW and 100 kW with the normal longitudinal emittance. The lowest RF voltage was set as 15 kV, 20 kV, and 30 kV for the cases of 30 kW, 50 kW, and 100 kW, respectively, while the flattop voltage of 156 kV remains the same for all the cases. The extraction moment was delayed to 20.20 ms in all these tests to facilitate the observation of the entire process of the bunch stretching and rotation.

The comparison between the desynchronization and synchronization on the measured bunch length for the three cases is shown in Fig. 14. One can clearly see that the desynchronization can help obtain shorter bunches than the synchronization in all the cases, as predicted by the theoretical expectation. In addition, one can observe that the influence of the desynchronization is much more significant on the shortest bunch length when the bunch emittance is larger or the lowest voltage is smaller. This is understandable because the bunch twisting during the rotation process is more critical with a longer bunch length just before the bunch rotation, which corresponds to larger longitudinal emittance and smaller lowest RF voltage.

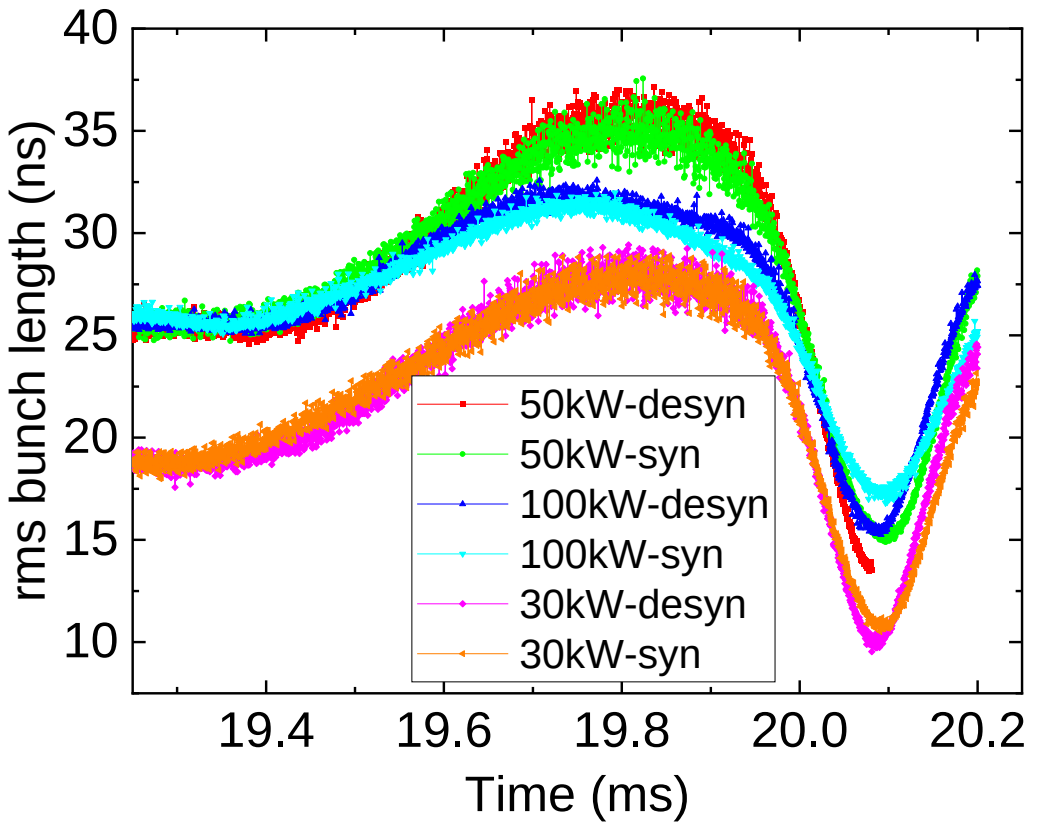

FIG. 14. Comparison between the desynchronization and synchronization on the measured bunch length under the equivalent beam power of 30 kW, 50 kW, and 100 kW. The non-adiabatic RF voltage risetime is 100 μs for all the cases. The extraction moment is 20.2 ms except 19.8 ms for the case of 50 kW with desynchronization.

## V. CONCLUSION AND OUTLOOK

Both the experimental studies (machine studies) and simulations show that the bunch rotation method can be practically applied to extract short bunches in a high-intensity rapid cycling proton synchrotron. As large as a factor of four in shortening the bunch length from the normal operation mode to a dedicated short bunch mode can be attained. The RF system based on ferrite-loaded cavities in the CSNS/RCS has been improved to allow special manipulations, such as the controlled desynchronization between the RF and the magnetic fields, fast phase shift, and rapid rising in voltage.

Different experimental settings in the CSNS/RCS were used to study the influence of beam intensity, longitudinal emittance, RF voltage decay time for adiabatic bunch stretching, RF voltage risetime for non-adiabatic bunch rotation, space charge, and the desynchronization, to the short bunch extraction. The experimental results are consistent with the simulations. The beam-shooting-target mode with the bunch compression was successfully operated for about 1 h at the beam power of 50 kW, which demonstrates that the bunch compression method can be applied in the operation of a practical high-power RCS. The bunch length of the extracted beam is 14.8 ns in rms or about 53% of the one in the 100-kW normal operation mode. The shortest extraction bunch was obtained with a reduced longitudinal emittance at injection and a reduced beam power of 30 kW, which gives about 26% of the one in the normal operation mode. The method of the controlled desynchronization between the RF and magnetic fields was successfully applied in the machine and confirmed its advantage in enhancing the bunch compression. Furthermore, the extraction bunch length was found modestly correlated with the adiabatic decay time and non-adiabatic risetime of the RF voltage. Space charge was found beneficial in obtaining shorter extracted bunches in the present machine conditions as its positive effect on the bunch lengthening process overtakes its detrimental impact on the bunch rotation process.

Despite the success, more efforts are still needed to make the method fully applicable to the daily operation of the accelerator, in particular on the further improvement of the RF system to work reliably at a small lowest voltage and fast voltage rising with heavy beam loading. In the future, CSNS will be upgraded to 500 kW mainly by increasing the injection energy from current 80 MeV to 300 MeV and adding a doubled harmonic RF system based on magnetic-alloy (MA) loaded RF cavities to compose a dual-harmonic acceleration [22]. On one hand, a larger longitudinal emittance coming with the dual-harmonic RF system limits the effect of bunch compression; on the other hand, the MA-loaded cavities have better performance in resisting the beam loading effect and the fast-rising of the RF voltage. Additionally, the MA-loaded cavities can work at the same RF frequency as the ferrite-loaded ones in the period of bunch compression to obtain higher flattop voltage to increase the ratio of the flattop to the lowest RF voltage.

The method developed here could benefit other similar facilities and future accelerator projects. For example, the future neutrino factory or muon collider demands a proton driver to deliver a proton beam of 4 MW and a few GeV with a very short bunch length. As pointed in the earlier Neutrino Factory study [23], when an RCS is chosen as the main acceleration stage of the proton driver, some kind of bunch compression is required. The bunch rotation in the RCS may avoid an additional bunch compressor ring.

## ACKNOWLEDGEMENTS

The authors would like to thank the CSNS colleagues for their strong support in the machine studies, especially W.L. Huang, R.R. Fan, W. Jiang, and Y. Li. The study is jointly supported by the National Key Research and Development Program (2016YFA0401601) and the National Natural Science Foundation of China (Projects 12035017, 11527811).